\documentclass[aps,prb,twocolumn,showpacs,superscriptaddress]{revtex4}
\usepackage{graphicx}
\usepackage{amsmath}

\begin{document}

%\preprint{}

\title{
Pinning Action of Correlated Disorder against Equilibrium Properties of
HgBa$_2$Ca$_2$Cu$_3$O$_x$: a Delicate Balance

}



\author{J. R. Thompson}
\affiliation{Oak Ridge National Laboratory, Oak Ridge, Tennessee,
             37831-6061 USA}
\affiliation{Department of Physics, University of Tennessee,
             Knoxville, Tennessee 37996-1200 USA}

\author{J. G. Ossandon}
\affiliation{Department of Engineering Sciences, University of Talca,
             Curico, Chile}

\author{L. Krusin-Elbaum}
\affiliation{IBM Watson Research Center, Yorktown, New York, 10598
             USA}

\author{D. K. Christen}
\affiliation{Oak Ridge National Laboratory, Oak Ridge, Tennessee,
             37831-6061 USA}

\author{H. J. Kim}
\altaffiliation{now at Department of Physics and Astronomy, 
Michigan State University, East Lansing, Michigan}
\affiliation{Department of Physics, University of Tennessee,
             Knoxville, Tennessee 37996-1200 USA}
\author{K. J. Song}

\altaffiliation{now at Korea Electrotechnology Research Institute,
Changwon, Kyung-Nam, 641-120 Korea}
\affiliation{Department of Physics, University of Tennessee,
             Knoxville, Tennessee 37996-1200 USA}

\author{K. D. Sorge}
\altaffiliation{now at Department of Physics, 
University of Nebraska, Lincoln, Nebraska}
\affiliation{Department of Physics, University of Tennessee,
             Knoxville, Tennessee 37996-1200 USA}

\author{J. L. Ullmann}
\affiliation{Los Alamos National Laboratory, Los Alamos, New Mexico,
             87545 USA}

%\thanks{}


\date{\today}

\begin{abstract}

We report significant alteration of the equilibrium properties of 
the superconductor HgBa$_2$Ca$_2$Cu$_3$O$_x$ when correlated
disorder in the form of randomly oriented columnar tracks 
is introduced via induced fission of Hg-nuclei. From studies of 
the equilibrium magnetization $M_{eq}$ and the persistent
current density over a wide range of temperatures, 
applied magnetic fields, and track densities up to 
a ``matching field'' of 3.4 Tesla, we observe that the
addition of more columnar tracks acting as pinning centers 
is progressively offset by reductions in the magnitude of
$M_{eq}$. Invoking anisotropy induced ``refocusing'' of the 
random track array and incorporating vortex-defect 
interactions, we find that this corresponds to
increases in the London penetration depth $\lambda$;  
this reduces the vortex line energy and consequently 
reduces the pinning effectiveness of the tracks.

\end{abstract}

%\keywords{}

\maketitle

%\section{}
%\subsection{}
%\subsubsection{}

\section{Introduction}

Much has been done over the past decade or so to understand the
interaction of superconducting vortices in high $T_c$ cuprates with
correlated disorder, which generally gives strong vortex pinning.
Most widely studied have been columnar defects, which typically are
formed by irradiation with energetic heavy ions; such particles are
highly ionizing and create tracks of amorphous material along their
path.
\cite{Civale92}
Many morphologies have been investigated, ranging from the familiar
parallel tracks to inclined defects, crossed arrays with ``designer
splay,'' etc.  Early theoretical insights and ideas came from Nelson
and Vinokur
\cite{NelsonVinokur}
and Hwa et al.
\cite{Hwa93}
Numerous experimental studies have shown that such defects are
supremely efficient in localizing vortices to their vicinity.  This is
manifest through strong enhancements of critical ($J_c$) and
persistent current densities in some regions of the phase space. It is
also clear is that outside these confined regions the localization
effects are lessened, in part by the complex vortex (variable range
hopping) dynamics
\cite{Blatter94}
which interferes and competes with pinning.

Vortex pinning by proton-generated fission defects was first
demonstrated in Bi-cuprate materials
\cite{LKE94,Safar95} and
can be generally applied in cuprate
superconductors that contain a sufficient density of heavy
nuclei.
\cite{Thompson97}
The method was devised to circumvent a problem of generating CD's
by heavy ion irradiation, namely their limited range, typically a few
tens of micrometers.  Krusin-Elbaum et al.
\cite{LKE94}
demonstrated an indirect formation of columnar defects by irradiation with deeply
penetrating 0.8~GeV protons.  An incident proton is
absorbed by a heavy nucleus (Hg in the present case) in the material, 
excites it, and induces it to fission 
into two particles with similar mass.  The recoiling fission 
fragments, each having $\sim$ 100~MeV
energy, form randomly oriented columnar defects
(CD's) deep within the superconductor.  In a different approach, 
Weinstein and coworkers 
created CD's by doping a superconductor with $^{235}$U, 
which is induced to fission by irradiation with thermal neutrons.
\cite{Schultz98}
High-$T_c$ materials containing randomly oriented CD's
exhibit a variety of interesting physical phenomena, e.g., a  
temperature-independent quantum tunneling of vortices in Bi-2212
materials.
\cite{Thompson99}
In studies of Hg-cuprates containing 1, 2, and 3 adjacent
CuO layers, it was shown
\cite{LKE98}
that sufficiently high superconductive anisotropy can lead to a
rescaling of the splayed landscape of random CD's.  As a consequence,
the pinning array and the applied field are ``refocused'' toward the
crystalline $c$-axis, even in polycrystalline materials. Later we 
use this feature for interpreting the present studies.

In addition to providing vortex pinning sites, correlated disorder 
doubtless also affects the equilibrium
superconducting properties, e.g., the characteristic length scales
such as $\lambda$ and $\xi$. What is little known, however, is how this
alters the pinning landscape; for example, an increase in $\lambda$
will reduce the vortex line energy and hence the pinning force will be
reduced.

In this work, we examine the equilibrium magnetization $M_{eq}$ in the mixed
state of the anisotropic cuprate HgBa$_2$Ca$_2$Cu$_3$O$_x$, both in
its virgin state and when containing randomly splayed columnar defects
installed by fission.  For virgin Hg-1223, the $M_{eq}$ can be scaled,
following a recent formulation.
\cite{LandauOtt02,LandauOtt03}
With the addition of splayed CD's, the magnitude of $M_{eq}$ decreases
significantly, especially at low fields. In addition, the defects
generate a pronounced deviation from the `standard' London field dependence,
giving an anomalous ``S'' shape to plots of $M_{eq}$ versus ln($H$). 
By using the concept that
sufficiently large anisotropy rescales the random track collection
into an equivalent parallel defect array,
\cite{LKE98}
we apply---in addition to a conventional London description---the
vortex-defect interaction model by Wahl et al.,
\cite{Wahl95}
which considers a Poisson distribution of parallel tracks.  We show that the
``S'' shape of $M_{eq}$ is accounted for by the magnetic coupling
between defects and vortices.  The London penetration depth
$\lambda$ significantly and progressively increases, in
accordance with the Wahl-Buzdin description.  This effect counteracts
the nominally positive influence of increasing the defect density
$B_{\Phi}$ and it plays a major role in determining the level 
of $B_{\Phi}$ that maximizes the current density $J_c$.

\section{Experimental Aspects}

Samples for study were bulk polycrystalline HgBa$_2$Ca$_2$Cu$_3$O$_x$
materials (Hg-1223) containing sets of 3 adjacent oxygen-copper
layers.  Small pieces, typically 30~mg mass and $\sim$ 1 millimeter
thickness, were all cut from the same pellet.  The samples were
irradiated at room temperature in air with 0.8~GeV protons at the Los
Alamos National Laboratory.  Proton fluences $\phi_p$ were 0, 1.0,
3.2, 10, 19, and $35 \times 10^{16}$~protons/cm$^2$, as determined
from the activation of Al dosimetry foils. The resulting density of
fission events is $N/V = \phi_{p} \sigma_{f} n_{\text{Hg}}$.  Here
$n_{\text{Hg}}$ is the number density of Hg nuclei and $\sigma_{f}
\simeq$ 80 millibarns is the cross-section for inducing a prompt
fission of Hg nuclei.  Each fission produces one CD, so $N/V$ is the
volume density of defects, also.  One can convert this into an
approximate area density of CD's by multiplying by the track length
$\sim$ 8~$\mu$m, and the area density can be reexpressed in units of a
matching field $B_{\Phi}$ by multiplying by the flux quantum
$\Phi_0$. Resulting values for the defect density are $B_{\Phi}$ = 0,
0.1, 0.3, 1.0, 1.9, and 3.4 Tesla, respectively. The fission process
is random in direction, giving randomly oriented CD's.  Transmission
electron microscopy (TEM) studies of both Bi-2212 and
Hg-cuprates
\cite{LKE97}
have demonstrated the presence of randomly oriented columnar defects
in these high-$T_c$ materials.

The superconductive properties of the virgin and processed samples
were investigated magnetically.  A SQUID-based magnetometer (model
Quantum Design MPMS-7), equipped with a high homogeneity 7~T
superconductive magnet, was used for studies in the temperature range
5--295~K, in applied fields to 6.5~T.  The superconductive transition
temperatures $T_c$ were measured in an applied field of 4~Oe (0.4~mT)
in zero-field-cooling (ZFC) and field-cooling (FC) modes.  The
resulting values for the onset temperature $T_c$ are 133, 134, 132.3,
132, 131, and 129.3~K, respectively.  Values for the Meissner (FC)
fraction $-4 \pi M/H$ lie in the range of 40--50\%, except at the
highest fluence where the fractional flux expulsion was 29\%.

The isothermal magnetization $M$ was measured as a function of applied
magnetic field.  Below $T_c$ and below the irreversibility line, the
magnetization was hysteretic due to the presence of intragrain
persistent currents.  From the magnetic irreversibility $\Delta M =
[M(H\downarrow)-M(H\uparrow)]$, the persistent current density $J$
was obtained using the Bean critical state relation $J \propto \Delta
M/r$, where $r \approx 4$~$\mu$m is the mean grain radius.
Measurements of the background magnetization in the normal state above
$T_c$ were used to correct the data in the superconducting state, in
order to obtain the equilibrium magnetization $M_{eq}$.  Above the
irreversibility line where $\Delta M = 0$, this process yields
$M_{eq}$ directly; near the irreversibility line where $\Delta M$ is
small, we obtain $M_{eq}$ from the (background-corrected) average
magnetization $[M(H\downarrow)+M(H\uparrow)]/2$, as illustrated
in the next section.

\section{Experimental Results}

\subsection{Equilibrium Properties of Virgin Hg-1223}

\begin{figure}
\includegraphics[width=8.5cm]{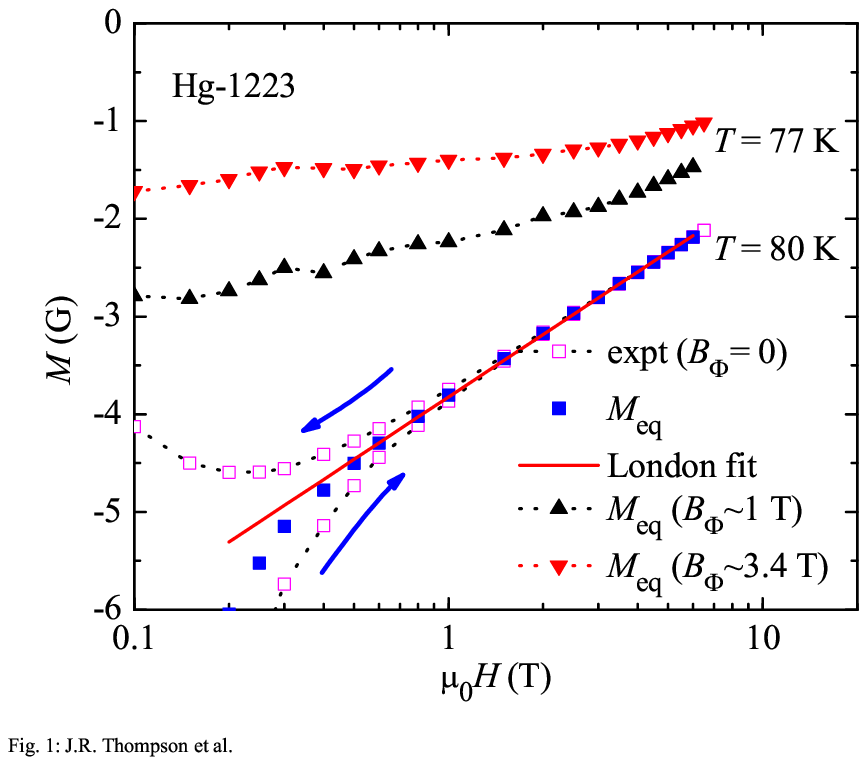}%
  \caption{\label{Fig1} (color online) The magnetization of virgin Hg-1223 at 80~K,
           versus applied magnetic field (log scale).  Open squares
           show the measured $M$ in increasing and decreasing field
           history, while closed squares show the average $M \approx
           M_{eq}$.  The solid line is a fit to the conventional
           London $\ln(H)$ dependence.  Data for Hg-1223 containing
           randomly oriented columnar defects are included for
           comparison.}
\end{figure}

We begin by considering the equilibrium magnetization in the mixed
state.  Figure
\ref{Fig1}
illustrates the method used to obtain $M_{eq}$ by plotting the
background-corrected experimental magnetization for the virgin
material (open squares) versus applied magnetic field $\mu_0 H$ on a
logarithmic scale.  As described above, the closed symbols show the
average $M$, which provides a very good approximation to $M_{eq}$ when
the hysteresis is small.  The solid line is a fit to conventional
London theory, which provides that $M_{eq}$ is directly proportional
to $(1/\lambda_{ab}^{2}) \times \ln(\beta B_{c2}/H)$, where
$\lambda_{ab}$ is the in-plane London penetration depth, $\beta
\approx 1.4$ is a constant
\cite{HaoClem}
of order unity, and $B_{c2}$ is the upper critical field.
\cite{Kogan88}
More explicitly, one has with $H \Vert$ c-axis of a uniaxial 
superconductor that
\begin{equation}
  \label{Eq1}
  M_{eq} = -(\Phi_0/32 \pi^2 \lambda_{ab}^2) \times 
\ln(\beta H_{c2}/H)
\end{equation}
If $H$ is oriented at angle $\theta$ with respect to the c-axis, the 
resulting response can be obtained from the anisotropy 
function\cite{Blatter92}
\begin{equation}
  \label{Eq2}
\varepsilon(\theta) = [\cos^2(\theta)+(1/\gamma^2)\sin^2(\theta)]^{1/2}
\end{equation}
where $\gamma = (m_c/m_{ab})^{1/2}$ is the mass anisotropy ratio.
For a randomly oriented polycrystalline material, the angularly 
averaged equilibrium magnetization is 
\begin{equation}
  \label{Eq3}
  \langle M_{eq} \rangle = -(\Phi_0/32 \pi^2 \lambda_{ab}^2) \times 
[(1/2) \ln(\beta H_{c2}/H) + (1/4)]
\end{equation}
in the limit that $\gamma$ is large.

The straight line fit in Fig.
\ref{Fig1}
shows that the London $\ln(H)$ logarithmic variation describes the
field dependence well.  Also shown for immediate qualitative
comparison are data for $M_{eq}$ of irradiated samples (at 77~K) with
defect densities $B_{\Phi} \approx 1.0$ and 3.4~T.

\begin{figure}
\includegraphics[width=8.5cm]  {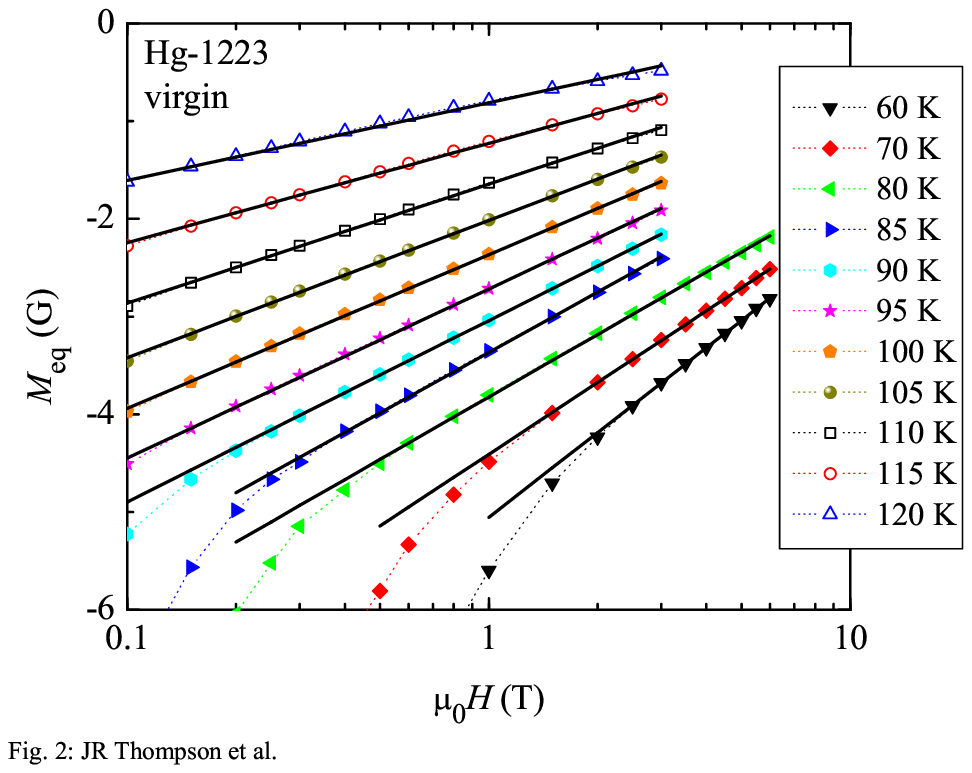}
  \caption{\label{Fig2} (color online) The equilibrium 
magnetization $M_{eq}(H,T)$
           plotted vs $\ln(H)$ for unirradiated Hg-1223 at the
           temperatures shown.  Lines are fits to conventional London
           theory.}
\end{figure}

Results for $M_{eq}$ in the virgin Hg-1223 material at other
temperatures are presented in Fig.
\ref{Fig2}.
One sees from the linear regression lines that the London field
dependence is followed over a wide range of fields and temperatures.
Deviations from linearity occur at low fields when $H$ approaches
$H_{c1}$ where simple London theory is not valid, and at low
temperatures where the materials are sufficiently hysteretic that the
averaging procedure illustrated in Fig.
\ref{Fig1}
is not valid. From Eq. \eqref{Eq3} and the 
slopes of the curves in Fig.
\ref{Fig2}, we obtain
values for the London penetration depth $\lambda_{ab}(T)$ in
the virgin material.  These results will be compared and contrasted
with values deduced for the irradiated Hg-1223. In addition, the
intercepts provide a measure of the upper critical field 
$e^{1/2} \beta B_{c2}(T)$, where $e \approx 2.718...$ is the natural logarithm base.  We now compare these results with 
those from a scaling analysis.

\begin{figure}
\includegraphics[width=8.5cm]  {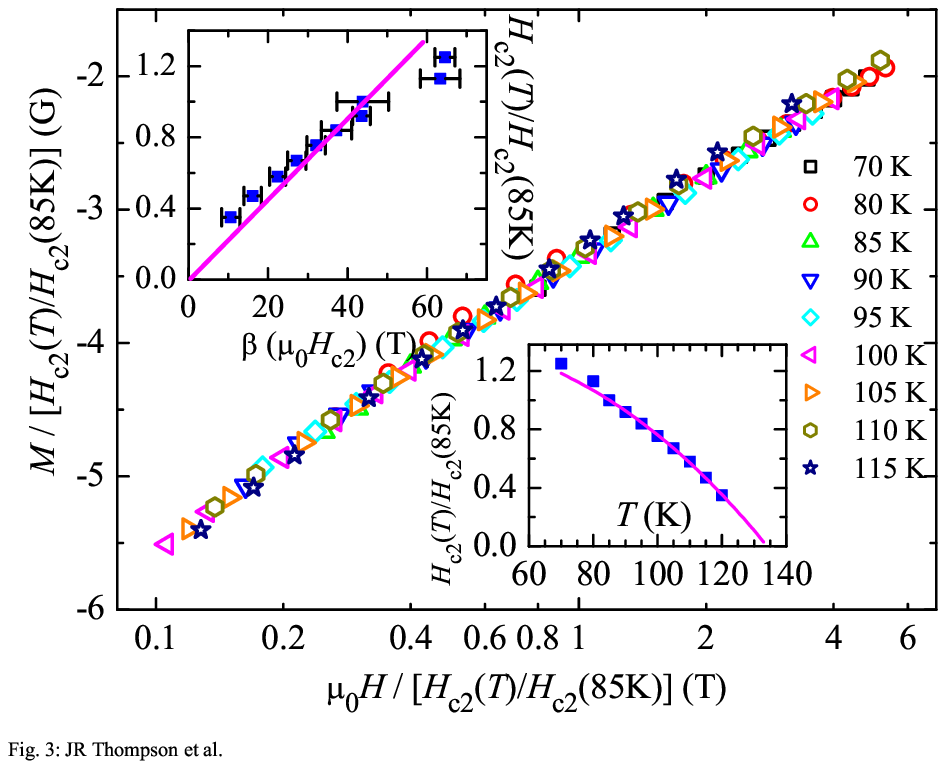}
  \caption{\label{Fig3} (color online) Scaling of the 
magnetization and applied field
           by the (relative) upper critical field $h_{c2}$ for virgin
           Hg-1223.  Lower inset shows scaling field $h_{c2}$ versus
           $T$, fitted to Eq.
           \eqref{Eq5}
           (solid line).  Upper inset compares values for $h_{c2}$
           with results for $\beta H_{c2}$ from London analysis in Fig
           \ref{Fig2}.}
\end{figure}

Recently, Landau and Ott
\cite{LandauOtt02}
demonstrated a useful scaling methodology based on
Ginzburg-Landau theory.  This has been applied to both single crystal
and aligned particle arrays
\cite{LandauOtt02}
and more recently to randomly oriented, polycrystalline
superconductors.
\cite{LandauOtt03}
If one assumes that the Ginzburg-Landau parameter $\kappa =
\lambda/\xi$ is independent of temperature, then $M_{eq}(H,T)$ is a
unique function of $H/H_{c2}(T)$.  Consequently it follows that the
equilibrium magnetizations in field $H$ at temperature $T$ and
(arbitrary) reference temperature $T_0$ are related as follows:
\begin{equation}
  \label{Eq4}
  M(H,T_0) = M(h_{c2}H,T)/h_{c2}
\end{equation}
where $h_{c2} = H_{c2}(T)/H_{c2}(T_0)$.  To apply this scaling
relationship, we chose $T_0 = 85$~K and manually varied $h_{c2}$ to
superposition the data for each temperature.  (While a term of the
form $c_0(T)H$ was included in the original analysis
\cite{LandauOtt02,LandauOtt03}
to account for additional contributions to $M$, we took a minimalist
approach of having $h_{c2}$ as the sole parameter.)  The results are
shown in Fig.
\ref{Fig3},
which plots $M/h_{c2}$ versus $H/h_{c2}$ for a wide range of
temperatures.  Given the simplicity of the approach, the scaling is
quite good.  The results for $h_{c2}(T)$, shown in the lower inset,
are well behaved.  Following Landau and Ott, we fit the temperature
dependence to an expression of the form
\begin{equation}
  \label{Eq5}
  h_{c2}(T) = h_{c2}(0)[1-(T/T_c)^\mu]
\end{equation}
for $T \ge 90$~K.  This describes the data well (solid curve in the
lower inset) and yields $\mu = 2.5$ and $T_c = 134$~K, which is quite
close to the diamagnetic onset of 133~K measured in low field.
Unfortunately, the scaling analysis provides only relative values for
$H_{c2}$.  So it is useful to compare these results with the
corresponding determinations of $\beta H_{c2}$ from the London
analysis.  This comparison is shown in the upper inset to Fig.
\ref{Fig3}.
The error bars show one standard deviation and the straight line
illustrates a simple direct proportionality.  While there seems to be a
systematic deviation from such a simple relation, most data lie within
two standard deviations of a linear dependence.

\subsection{Equilibrium Properties of Hg-1223 with Randomly Oriented CDs}

We now address the question of how the addition of randomly oriented
columnar defects modifies the equilibrium magnetization in the mixed
state.  The qualitative effects of the CD's are illustrated in Fig.
\ref{Fig1},
which includes data for $M_{eq}$ at 77~K for irradiated samples with
defect densities $B_{\Phi} \approx 1.0$ and 3.4~T.  Comparing these
data with the virgin curve shows that the CD's reduce considerably 
the magnitude of
the equilibrium magnetization, especially at lower
fields.  In addition, they generate a pronounced deviation from the
``standard'' London field dependence, giving $M_{eq}$ an ``S''-like
dependence on field.  This ``anomalous'' behavior has been observed
previously in cuprates containing \emph{parallel} columnar defects
formed by 5.8~GeV Pb-ions, in thallium-based single crystals
\cite{Wahl95};
in Bi-2223 tapes
\cite{Li96};
and in Bi-2212 single crystals.
\cite{vanderBeek96,vanderBeek00}
    
These changes in the equilibrium magnetization have been attributed to
magnetic interactions between the vortex lattice and the columnar
defects.  By occupying a pinning site, a vortex gains pinning energy.
This reduction in system energy must exceed the energy increase
arising from direct intervortex repulsion when a vortex is displaced
from its natural position in the lattice to a particular columnar
defect. For a defect geometry with parallel tracks that have a Poisson
distribution of separations, Wahl et al.
\cite{Wahl95}
obtained an expression for $M_{eq}$ that describes reasonably well the
``S''-shaped field dependence in Tl-cuprate crystals containing
parallel CD's.

\begin{figure}
  \includegraphics[width=8.5cm]   {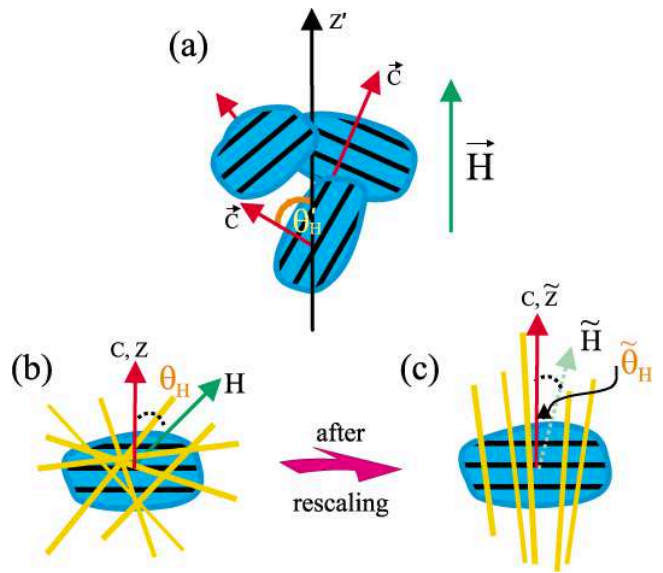}
  \caption{\label{Fig4} (color online) An illustration  
of the refocussing process.
           (a) Polycrystalline material with randomly distributed
           $c$-axes.  (b) One grain with randomly oriented columnar
           defects, in magnetic field with arbitrary orientation
           $\theta_{H}$.  (c) after rescaling, both the effective
           field and defect array are tilted close to the
           $c$-axis.
           \cite{LKE98}}
\end{figure}

In the Hg-cuprates investigated here, one might expect that the
randomly oriented columnar microstructure should entangle the
vortices.  Consequently it is somewhat curious that the $M_{eq}$ in
polycrystalline materials with random CD's can resemble single
crystals with parallel defects.  The operative 'refocussing' 
process is sketched in
Fig.
\ref{Fig4}.
(a) In real space, the sample consists of grains randomly oriented
with respect to the applied field $H$.  (b) After irradiation, there
are CD's randomly oriented about the $c$-axis, which makes some angle
$\theta_H$ with respect to an applied magnetic field.  Now, for highly
anisotropic single crystals, it is long recognized that only the
normal component of field is effective.
\cite{Blatter92}
More generally, rescaling gives a field $\tilde{H}$ lying near the
$c$-axis; similar arguments
\cite{LKE98}
lead to a ``refocussing'' of the CD's into a narrow distribution about
the $c$-axis.  According to the theoretical development,
\cite{LKE98}
the complexity of randomly oriented columnar defects in a
polycrystalline material is reduced to a considerable degree by a
large superconducting anisotropy, restoring the simple physics of a
crystal with parallel pins.  Indeed, studies of polycrystalline
Hg-cuprates demonstrated a recovery of vortex variable range hopping
(VRH), very similar to that observed in YBaCuO single crystals
containing parallel CD's.
\cite{ThompsonPRL97}
Among the Hg-cuprates with 1, 2, and 3 adjacent Cu-O layers, the
recovery of VRH was most pronounced in the Hg-1223 material with the
largest mass anisotropy parameter $\gamma$.  We have suggested
\cite{Ossandon01}
that this scenario---an anisotropy-induced ``refocussing'' of the
defects and field toward the crystalline $c$-axis---can account for
the ``S''-like dependence of $M_{eq}$ on field in highly
anisotropic Tl-2212 materials, where $\gamma \sim 200$.  Here we test
and confirm the validity of the rescaling formalism in these Hg-1223
materials with significantly lower $\gamma \sim 60$.  Below 
we show that vortex pinning by
random CD's provides a reasonable qualitative explanation for the
reduction in equilibrium magnetization, despite the complexity of the
materials in real space.

Let us now use the vortex-defect interaction theory of Wahl et al.
\cite{Wahl95}
to model our experimental data.  From the theory, one has for $H \Vert$
c-axis that
\begin{eqnarray}
% \label{Eq6}
  M_{eq} & = & -(\varepsilon_0/2\Phi_0)\times
                   \ln(\beta H_{c2}/B) \nonumber\\
           & & +\frac{U_0}{\Phi_0}
                \left\{
                  1 - \left[
                        1+\frac{U_0B_{\Phi}}{\varepsilon_0 B}
                      \right]
                  \exp\left(
                        -\frac{U_0B_{\Phi}}{\varepsilon_0 B}
                      \right)
                \right\}
\end{eqnarray}
where $\varepsilon_0 = [\Phi_0 /4 \pi \lambda_{ab}]^2$ is the line
energy, $U_0 \propto \varepsilon_0$ is the pinning energy, 
and $B = (H + 4 \pi M) = H$ since
$M$ is small.  The first term is the conventional London expression.
The second is an added term to account for interactions; it is most
significant in intermediate fields and it vanishes in large fields $B
\gg B_{\Phi}$, where vortices greatly outnumber defects.  
For $H$ inclined at angle $\theta$ from the c-axis, one can introduce 
the angular dependence of the line energy, pinning energy, and 
$B_{c2}$ via the anisotropy function Eq. \eqref{Eq2}.  
An angular average for a randomly oriented polycrystalline 
material leads to the following:

\begin{eqnarray}
% \label{Eq7}
 \langle M_{eq} \rangle & = & -(\varepsilon_0/2\Phi_0)\times
              [(1/2)\ln(\beta H_{c2}/B)+(1/4)] \nonumber\\
           & & +\frac{U_0}{2\Phi_0}
                \left\{
                  1 - \left[
                        1+\frac{U_0B_{\Phi}}{\varepsilon_0 B}
                      \right]
                  \exp\left(
                        -\frac{U_0B_{\Phi}}{\varepsilon_0 B}
                      \right)
                \right\}
\end{eqnarray}
Here the symbols have the same meaning as above and 
we assume that $\gamma \gg 1$.

\begin{figure}
\includegraphics[width=8.5cm]  {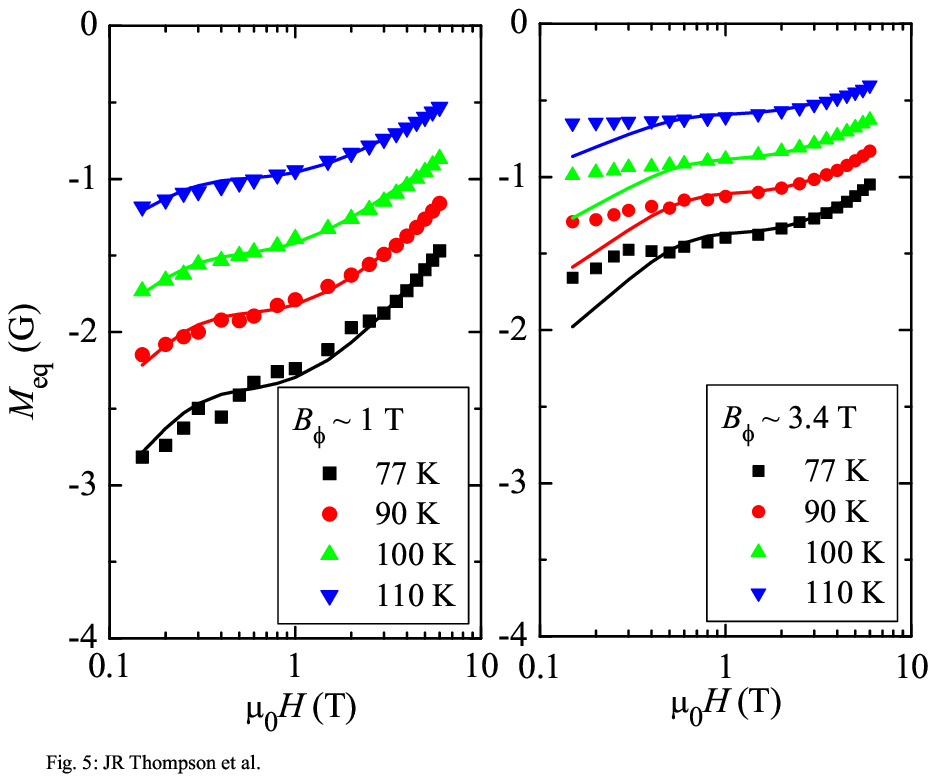}
  \caption{\label{Fig5} (color online)  Plots of $M_{eq}(H,T)$ 
vs. $\ln(H)$ for
           irradiated Hg-1223 with (a) $B_{\Phi} \sim 1$~T and (b)
           with $B_{\Phi} \sim 3.4$~T.  Symbols are experimental
           results, while lines show modeling of data using Eq. 7
           %\eqref{Eq7}
           where effects of vortex-defect interactions are included.}
\end{figure}

In using this expression, Eq. 7, to model the data at various 
temperatures $T$ (77,
90, 100, and 110~K), our objective was to maintain an internally
consistent set of parameters.  The results are shown as solid lines in
Fig.
\ref{Fig5}a
for the sample with $B_{\Phi} \sim 1$~T and in Fig.
\ref{Fig5}b
for the sample with a higher defect density.  Given the complexity of
the polycrystalline material and of the vortex and defect arrays, the
description of the experimental data (filled symbols) is relatively
good.  The values for the pinning energy are a significant fraction of
the line energy and are quite reasonable, with $U_0 = 0.8
\varepsilon_0$ for both materials.  To obtain
reasonable modeling of the data for the two irradiated materials, we
used values of 1.4 and 2.8~T, respectively, for the effective defect
densities $B_{\Phi}$.  The differences from the nominal values
(calculated from the proton fluences) may arise from some overlap of
tracks at the highest fluence combined with uncertainty in calculating
the production rate for CD's.  
% It is also possible that the effective
% values for $B_{\Phi}$ compensate for other factors, such as residual
% entanglement and angular distribution in the CD array.  Indeed, for
In analyzing the more highly anisotropic Tl-2212 materials 
($\gamma \sim 200$), comparable modeling
\cite{Ossandon01}
was achieved with $B_{\Phi}$ values varying by $\sim$ 20\% 
from those calculated from the proton fluence. 

\begin{figure}
\includegraphics[width=8.5cm]  {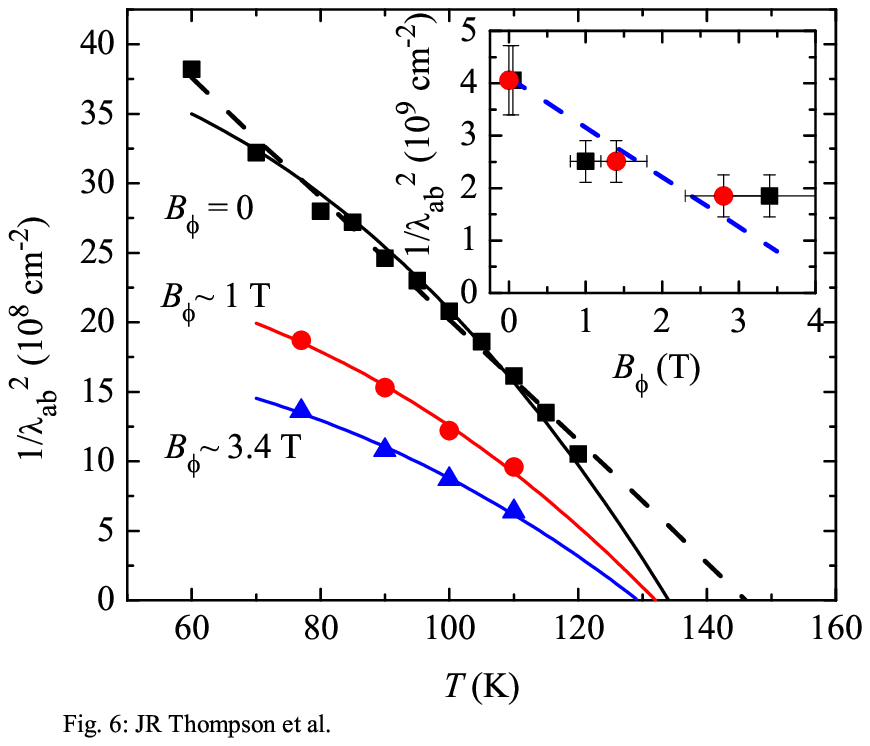}
  \caption{\label{Fig6} (color online)  The temperature 
dependence of the London
           penetration depth 1/$\lambda_{ab}^2$.  Values for the virgin
           sample come from a conventional London analysis of the data
           in Fig.
           \ref{Fig5};
           for the irradiated materials, values come from modeling the
           equilibrium magnetization shown in Fig.
           \ref{Fig6}.
           Inset shows values for 1/$\lambda_{ab}^2$, extrapolated to
           $T=0$, plotted versus defect density; square symbols denote
           $B_{\Phi}$ values calculated from the proton fluence and
           circular symbols are values used in the modeling.}
\end{figure}

An important feature of any superconducting material is its London penetration
depth.  Figure
\ref{Fig6}
summarizes the results for these materials in a plot of
1/$\lambda_{ab}^2$ versus temperature.  Values for the virgin sample
were obtained from a standard London analysis of the data in Fig.
\ref{Fig2}.
An extrapolation of a Ginzburg-Landau linear dependence near $T_c$
(dashed line) to $T = 0$ yields a value $\lambda_{ab}(T=0) = 174$~nm, 
but the implied $T_c$-value is too high.
Alternatively, we can invoke the underlying assumption of the scaling
procedure
\cite{LandauOtt02}
that $\kappa$ is temperature independent, which means that
$1/\lambda_{ab}^2$ should have the same temperature dependence as $h_{c2}$.
Hence we fit the present data for the virgin sample ($T \ge 90$~K)
with the same temperature dependence in Eq.
\eqref{Eq5}
using the same values for $T_c$ and $\mu$ and varying only the
prefactor $1/\lambda^2(T=0)$, obtaining $\lambda_{ab}(T=0) = 157$~nm,
which is comparable with earlier determinations.
\cite{ThompsonReview98}.  
The resulting fit (solid line in main
figure) is `natural,' as illustrated by the fact that its
extrapolation to lower $T$ lies very near the lower temperature data.

In Fig.
\ref{Fig6},
the $\lambda_{ab}$-values for the materials with CD's were obtained 
from the modeling
procedure. All of these results are reasonably behaved, with
1/$\lambda_{ab}^2$ varying roughly linearly with $T$ at high
temperature, but extrapolating to $T_c$-values that are rather high.
Thus we also fit the data for the irradiated samples with the temperature
dependence in Eq.
\eqref{Eq5},
again setting $\mu$ = 2.5 and fixing $T_c$ at the experimental values
measured in low field, so that only the prefactor $1/\lambda^2(T=0)$ 
was varied.  The resulting fits, which are shown as solid lines in 
the main figure, describe the data quite well.

It is clear that proton irradiation increased $\lambda_{ab}$ significantly.
This is expected, from two theoretical perspectives. First,
conventional GLAG (Ginzburg-Landau-Abrikosov-Gorkov) theory
\cite{Orlando79}
predicts that the penetration depth $\lambda^2$ increases as
$(1+\xi_0/\ell)$ when the electronic mean free path $\ell$ is reduced
by scattering, as produced by irradiation-generated defects from
neutrons and secondary protons released by spallation.  Second, the
theory of Wahl-Buzdin
\cite{Wahl95}
provides that introducing CD's increases the penetration depth as
\begin{equation}
  \label{Eq8}
  \lambda^{-2}(B_{\Phi}) = \lambda^{-2}(B_{\Phi} 
                         = 0) \times [1-2 \pi R^2 B_{\Phi}/\Phi_0]
\end{equation}
where $R$ is the radius of a columnar track.  The combination of
these two effects leads to the significant, progressive increases in
$\lambda$ observed in Fig.
\ref{Fig6}.
To compare with Eq.
\eqref{Eq8},
the inset of Fig.
\ref{Fig6}
shows 1/$\lambda_{ab}^2(T=0)$ (the prefactors from fitting Eq.
\eqref{Eq5}) plotted versus defect density.  
These are plotted as a function of $B_{\Phi}$-values obtained both
from the proton fluence (squares) and the modeling 
procedure (circles), as discussed above.  
The straight line in the inset illustrates the theoretical
dependence in Eq.
\eqref{Eq8}
and agrees qualitatively with the data; the resulting value 
$R \approx 8.7$~nm is somewhat smaller than the 11~nm value
obtained for Tl-2212 materials.
\cite{Ossandon01}
In each case, it is likely that the CD's have a large effective 
radius due to oblique passage of ions through CuO planes and 
an oxygen-depleted region surrounding the amorphous track.
\cite{Zhu93}

An interesting consequence of a larger London penetration depth is
that the vortex line energy $\varepsilon_0 \sim 1/\lambda_{ab} ^2$ 
decreases significantly.
As a result, one can expect the vortex pinning energy of a CD and its
pinning force to decrease, reducing its effectiveness.  A progressive
decrease of $\varepsilon_0$ can be expected to counteract the
nominally positive influence of increasing the defect density
$B_{\Phi}$.  Consequently it is interesting and useful to examine how
the irreversible properties of the materials change with defect
density, which we now consider.

\subsection{Irreversible Properties and Persistent Current Density}

\begin{figure}
%   \includegraphics[width=3.3in,
% %                    height=5.5in ,
%                    keepaspectratio=true,
%                    bb=llxin llyin urxin uryin,
%                    clip=true]
\includegraphics[width=8.5cm, keepaspectratio=true] 
 {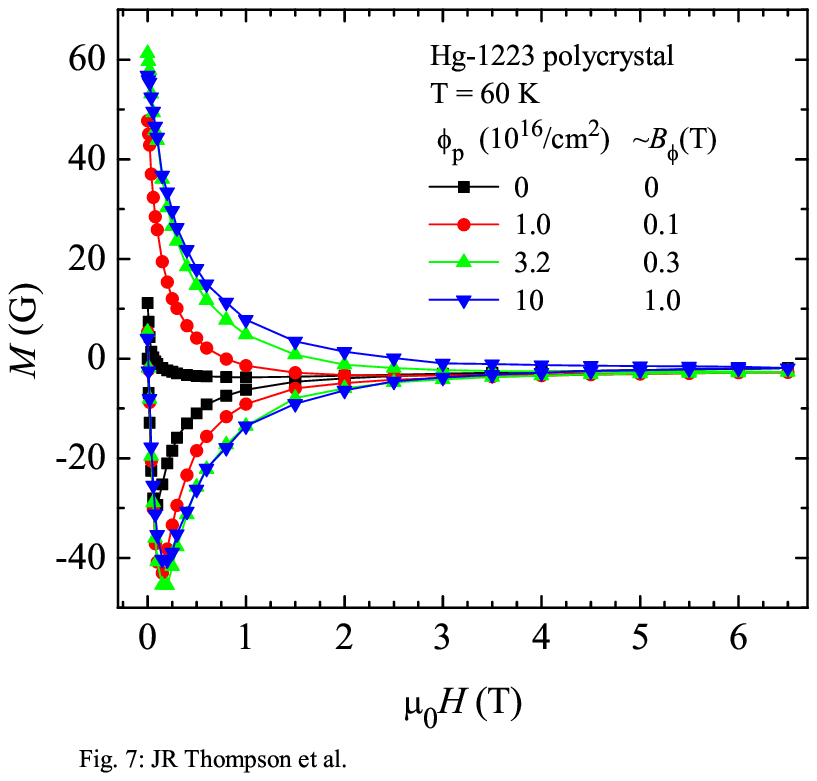}
  \caption{\label{Fig7} (color online) The magnetization of polycrystalline Hg-1223
           materials at 60~K, versus applied magnetic field.  Samples
           were irradiated with various fluences $\phi_p$ of 0.8~GeV
           protons, as shown, to create randomly oriented columnar
           defects with approximate ``matching fields'' $B_{\Phi}$.}
\end{figure}

As is well recognized, the addition of correlated disorder can
increase the pinning of vortices, often quite significantly.  This is
illustrated in Fig.
\ref{Fig7},
which shows the magnetization $M(H)$ at 60~K vs magnetizing field $H$,
for various defect densities.  With increasing fluence, the
``hysteresis loops'' increase in width and become symmetric about the
$M = 0$ axis.

For the virgin sample, the asymmetry, in combination with the fact
that the decreasing field branch lies near the $M = 0$ axis, suggests
that surface barriers
\cite{Sun94,Lewis95,Kim95} 
contribute to the observed hysteresis in this case.  It is interesting
to note that a relatively low dosage of CD's, $B_{\Phi} = 0.1$~T,
symmetrizes the $M(H)$ loop significantly; in particular, the
magnitude of $M$ in the decreasing field branch is much larger.  These
features imply that the addition of CD's suppresses the surface
barrier, as recently discussed by Koshelev and Vinokur.
\cite{Koshelev01}

\begin{figure}
\includegraphics[width=8.5cm]   {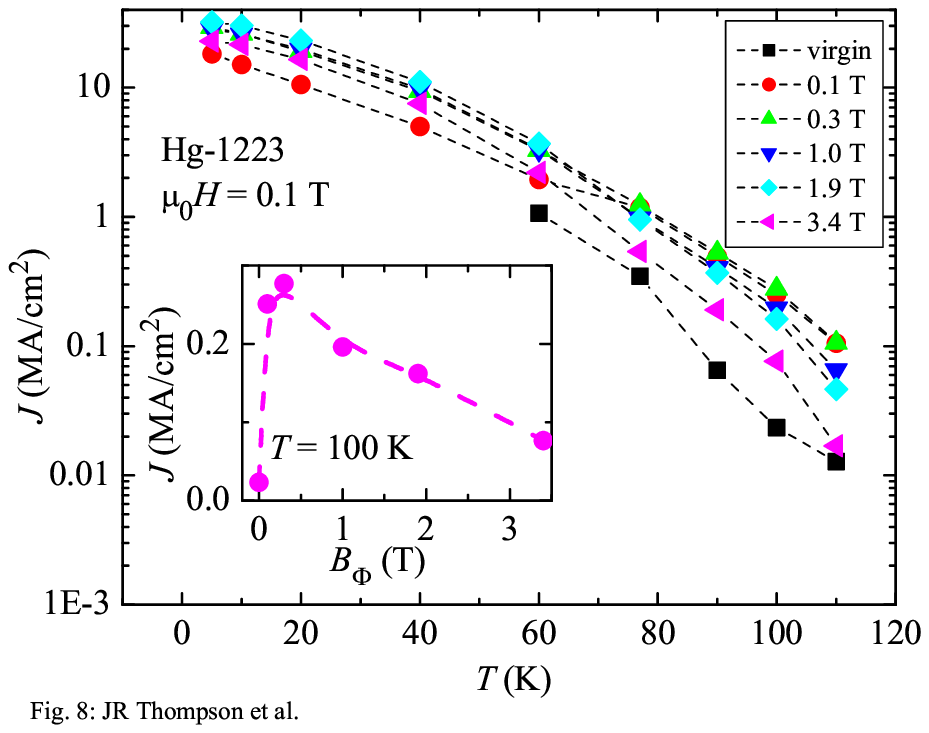}
  \caption{\label{Fig8} (color online)  The intra-granular 
persistent current density
           $J$ vs temperature for various irradiated materials,
           measured in applied magnetic field of 0.1~T.  The inset
           shows that $J$ at 100~K is maximized at a defect density of
           0.1--0.3~T.}
\end{figure}

\begin{figure}
\includegraphics[width=8.5cm]   {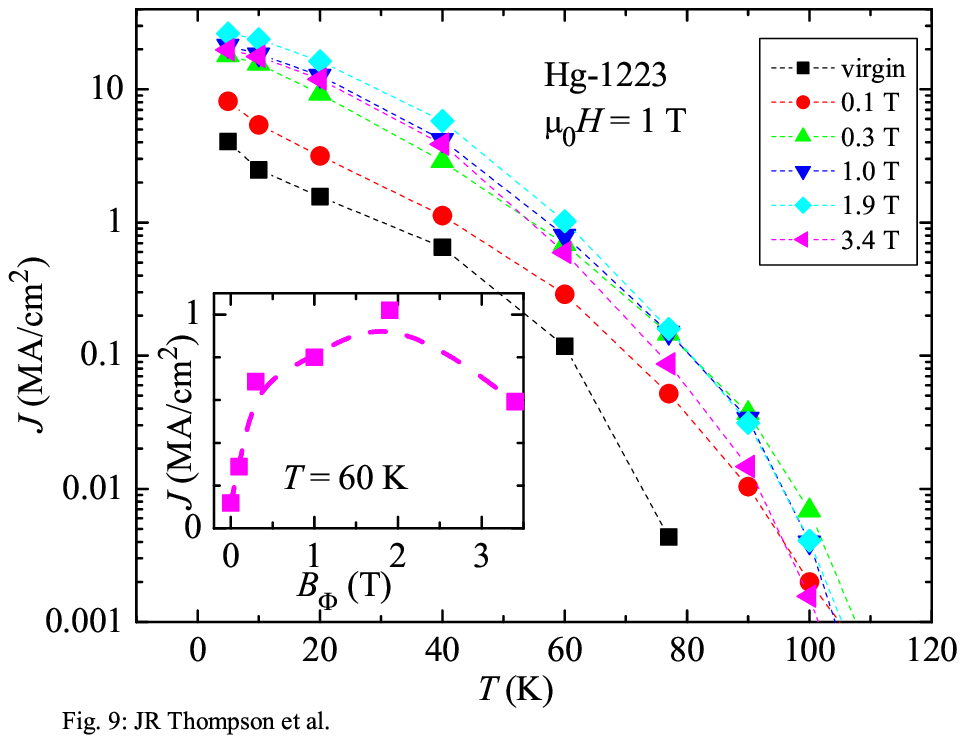}
  \caption{\label{Fig9} (color online)  The intra-granular 
persistent current density
           $J$ vs temperature for various irradiated materials,
           measured in applied magnetic field of 1~T.  
Inset: $J$ at
           60~K versus defect 
density $B_{\Phi}$ exhibits a maximum at
           a defect density of $\sim2$~T.}
\end{figure}

Using the Bean model, one may obtain the persistent current density
$J(H,T)$.  For these weakly linked polycrystalline materials, the
magnetization reflects the \emph{intragranular} current density.  Some
results of this analysis are shown in Fig.
\ref{Fig8}
and Fig.
\ref{Fig9}
as plots of $J$ versus temperature $T$, in applied fields of 0.1 and 1
Tesla, respectively.  The enhancement in $J$ is modest in low fields;
this can be attributed to the presence in the virgin sample of both
naturally occurring defects that provide some pinning and the likely
influence of surface barrier effects, as already noted. In higher
fields, the contribution of the random CD's is more apparent and $J$
is enhanced by about an order-of-magnitude.  For high-$T_c$ materials
with these angularly dispersed defects, one generally achieves the
maximum $J$ at some defect density $B_{\Phi}$ near 0.2--2~T.  This
optimization is shown by the insets in Figs.
\ref{Fig8}
and
\ref{Fig9}.
For an applied field of 0.1~T (Fig.
\ref{Fig8}
inset), $J$ at 100~K is largest for $B_{\Phi} = 0.1$--0.3~T.  For the
second example with a field of 1~T (Fig.
\ref{Fig9}
inset), $J$ at 60~K is largest for $B_{\Phi}$ near 2~T.
Qualitatively, the optimum defect density in each case is comparable
with the vortex density, i.e., about $2\times$ larger.

% fig 8 and 9 were here.

For higher defect densities, the current density $J$ decreases.  This
is due primarily to a suppression of the vortex line energy and
pinning energy, as a consequence of the large, experimentally observed 
increases in
the London penetration depth.  Thus we believe that that \emph{the
competition between increasing defect density and weakening line
energy play a major role in determining the range of $B_{\Phi}$ that
maximizes the current density} $J$, as illustrated in the insets of
Figs.
\ref{Fig8}
and
\ref{Fig9}.
This competition is reminiscent of the formation of point-like defects
in YBa$_2$Cu$_3$O$_{7-\delta}$, where progressive removal of oxygen
creates defects, but also reduces their effectiveness through
increases in the superconductive length scales.
\cite{Ossandon92a,Ossandon92b}

An additional mechanism for reducing $J$ at higher defect densities
may be that the presence of additional CD's helps to initiate the
hopping of vortices to nearby empty pinning sites.  A third potential
influence is a suppression of $T_c$; however, the change in $T_c$ is
small and these measurements are far from $T_c$, so this contribution
is expected to be small.

\section{Summary}

We irradiated polycrystalline HgBa$_2$Ca$_2$Cu$_3$O$_x$ materials with
0.8~GeV protons to produce randomly oriented columnar defects via a
fission process.  Adding random linear defects significantly reduced
the magnitude of the equilibrium magnetization $M_{eq}$ and changed
its dependence on magnetic field from a simple London $\ln(H)$ form
(as observed for the virgin material) to a more complex ``S''-shaped
dependence.  For the virgin superconductor, $M_{eq}(H,T)$ can be
scaled by the relative upper critical field $h_{c2}(T)$, as recently
discussed;
\cite{LandauOtt03}
these results compare reasonably with values from a London analysis.
For materials with random CD's, we invoke an anisotropy-induced
``refocusing'' of the vortex-defect array and model these results
using the theory of Wahl-Buzdin that incorporates vortex-defect
interactions.  This analysis shows that the addition of random
columnar defects increases the London penetration depth markedly.  The
success of this analysis demonstrates the essential correctness and
applicability of the refocussing mechanism in these materials with
intermediate anisotropy $\gamma \sim 60$.  Enhancements in vortex
pinning increased the persistent current density $J$, with the optimum
defect density depending on the range of field and temperature.
However, a high density of CD's increases $\lambda$ and decreases the
line energy, which diminishes their effectiveness for pinning
vortices.  Finally, the study provides qualitative evidence for the
suppresion of surface barriers in these Hg-1223 materials by a low
density of CD's.  Overall, we find a delicate balance between increases 
in the density of correlated disorder -- randomly oriented columnar
defects -- and decreases in the equilibrium superconductive properties, 
which progressively reduce the effectiveness of each defect added.

\begin{acknowledgments}
We thank M. Paranthaman for providing the starting Hg-1223 materials
used in this study, and thank I. L. Landau for communicating their
results prior to publication.  The work of JGO was supported in part
by the Chilean FONDECYT, grant \# 1000394.  Oak Ridge National
Laboratory is managed by UT-Battelle, LLC for the U.S. Department of
Energy under contract DE-AC05-00OR22725. Los Alamos National
Laboratory is funded by the US Department of Energy under contract
W-7405-ENG-36.
\end{acknowledgments}


\begin{thebibliography}{31}
\expandafter\ifx\csname natexlab\endcsname\relax\def\natexlab#1{#1}\fi
\expandafter\ifx\csname bibnamefont\endcsname\relax
  \def\bibnamefont#1{#1}\fi
\expandafter\ifx\csname bibfnamefont\endcsname\relax
  \def\bibfnamefont#1{#1}\fi
\expandafter\ifx\csname citenamefont\endcsname\relax
  \def\citenamefont#1{#1}\fi
\expandafter\ifx\csname url\endcsname\relax
  \def\url#1{\texttt{#1}}\fi
\expandafter\ifx\csname urlprefix\endcsname\relax\def\urlprefix{URL }\fi
\providecommand{\bibinfo}[2]{#2}
\providecommand{\eprint}[2][]{\url{#2}}

\bibitem[{\citenamefont{Civale et~al.}(1992)\citenamefont{Civale, Worthington,
  Krusin-{E}lbaum, Marwick, Holtzberg, Thompson, Kirk, and Wheeler}}]{Civale92}
\bibinfo{author}{\bibfnamefont{L.}~\bibnamefont{Civale}},
  \bibinfo{author}{\bibfnamefont{T.~K.} \bibnamefont{Worthington}},
  \bibinfo{author}{\bibfnamefont{L.}~\bibnamefont{Krusin-{E}lbaum}},
  \bibinfo{author}{\bibfnamefont{A.~D.} \bibnamefont{Marwick}},
  \bibinfo{author}{\bibfnamefont{F.}~\bibnamefont{Holtzberg}},
  \bibinfo{author}{\bibfnamefont{J.~R.} \bibnamefont{Thompson}},
  \bibinfo{author}{\bibfnamefont{M.~A.} \bibnamefont{Kirk}}, \bibnamefont{and}
  \bibinfo{author}{\bibfnamefont{R.}~\bibnamefont{Wheeler}},
  \bibinfo{journal}{J. Metals} \textbf{\bibinfo{volume}{44}},
  \bibinfo{pages}{60} (\bibinfo{year}{1992}).

\bibitem[{\citenamefont{Nelson and Vinokur}(1993)}]{NelsonVinokur}
\bibinfo{author}{\bibfnamefont{D.~R.} \bibnamefont{Nelson}} \bibnamefont{and}
  \bibinfo{author}{\bibfnamefont{V.~M.} \bibnamefont{Vinokur}},
  \bibinfo{journal}{Phys. Rev. B} \textbf{\bibinfo{volume}{48}},
  \bibinfo{pages}{13060} (\bibinfo{year}{1993}).

\bibitem[{\citenamefont{Hwa et~al.}(1993)\citenamefont{Hwa, {Le Doussal},
  Nelson, and Vinokur}}]{Hwa93}
\bibinfo{author}{\bibfnamefont{T.}~\bibnamefont{Hwa}},
  \bibinfo{author}{\bibfnamefont{P.}~\bibnamefont{{Le Doussal}}},
  \bibinfo{author}{\bibfnamefont{D.~R.} \bibnamefont{Nelson}},
  \bibnamefont{and} \bibinfo{author}{\bibfnamefont{V.~M.}
  \bibnamefont{Vinokur}}, \bibinfo{journal}{Phys. Rev. Lett.}
  \textbf{\bibinfo{volume}{71}}, \bibinfo{pages}{3545} (\bibinfo{year}{1993}).

\bibitem[{\citenamefont{Blatter et~al.}(1994)\citenamefont{Blatter, Feigel'man,
  Geshkenbein, Larkin, and Vinokur}}]{Blatter94}
\bibinfo{author}{\bibfnamefont{G.}~\bibnamefont{Blatter}},
  \bibinfo{author}{\bibfnamefont{M.~V.} \bibnamefont{Feigel'man}},
  \bibinfo{author}{\bibfnamefont{V.~B.} \bibnamefont{Geshkenbein}},
  \bibinfo{author}{\bibfnamefont{A.~I.} \bibnamefont{Larkin}},
  \bibnamefont{and} \bibinfo{author}{\bibfnamefont{V.~M.}
  \bibnamefont{Vinokur}}, \bibinfo{journal}{Rev. Mod. Phys.}
  \textbf{\bibinfo{volume}{66}}, \bibinfo{pages}{1125} (\bibinfo{year}{1994}).

\bibitem[{\citenamefont{Krusin-{E}lbaum
  et~al.}(1994)\citenamefont{Krusin-{E}lbaum, Thompson, Wheeler, Marwick, Li,
  Patel, Shaw, Lisowski, and Ullmann}}]{LKE94}
\bibinfo{author}{\bibfnamefont{L.}~\bibnamefont{Krusin-{E}lbaum}},
  \bibinfo{author}{\bibfnamefont{J.~R.} \bibnamefont{Thompson}},
  \bibinfo{author}{\bibfnamefont{R.}~\bibnamefont{Wheeler}},
  \bibinfo{author}{\bibfnamefont{A.~D.} \bibnamefont{Marwick}},
  \bibinfo{author}{\bibfnamefont{C.}~\bibnamefont{Li}},
  \bibinfo{author}{\bibfnamefont{S.}~\bibnamefont{Patel}},
  \bibinfo{author}{\bibfnamefont{D.~T.} \bibnamefont{Shaw}},
  \bibinfo{author}{\bibfnamefont{P.}~\bibnamefont{Lisowski}}, \bibnamefont{and}
  \bibinfo{author}{\bibfnamefont{J.}~\bibnamefont{Ullmann}},
  \bibinfo{journal}{Appl. Phys. Lett.} \textbf{\bibinfo{volume}{64}},
  \bibinfo{pages}{3331} (\bibinfo{year}{1994}).

\bibitem[{\citenamefont{Safar et~al.}(1995)\citenamefont{Safar, Cho, Fleshler,
  Maley, Willis, Coulter, Ullmann, Lisowski, Riley, Rupich et~al.}}]{Safar95}
\bibinfo{author}{\bibfnamefont{H.}~\bibnamefont{Safar}},
  \bibinfo{author}{\bibfnamefont{J.~H.} \bibnamefont{Cho}},
  \bibinfo{author}{\bibfnamefont{S.}~\bibnamefont{Fleshler}},
  \bibinfo{author}{\bibfnamefont{M.~P.} \bibnamefont{Maley}},
  \bibinfo{author}{\bibfnamefont{J.~O.} \bibnamefont{Willis}},
  \bibinfo{author}{\bibfnamefont{J.~Y.} \bibnamefont{Coulter}},
  \bibinfo{author}{\bibfnamefont{J.~L.} \bibnamefont{Ullmann}},
  \bibinfo{author}{\bibfnamefont{P.~W.} \bibnamefont{Lisowski}},
  \bibinfo{author}{\bibfnamefont{G.~N.} \bibnamefont{Riley}},
  \bibinfo{author}{\bibfnamefont{M.~W.} \bibnamefont{Rupich}},
  \bibnamefont{et~al.}, \bibinfo{journal}{Appl. Phys. Lett.}
  \textbf{\bibinfo{volume}{67}}, \bibinfo{pages}{130} (\bibinfo{year}{1995}).

\bibitem[{\citenamefont{Thompson
  et~al.}(1997{\natexlab{a}})\citenamefont{Thompson, Krusin-{E}lbaum, Christen,
  Song, Paranthaman, Ullmann, Wu, Ren, Wang, Tkaczyk et~al.}}]{Thompson97}
\bibinfo{author}{\bibfnamefont{J.~R.} \bibnamefont{Thompson}},
  \bibinfo{author}{\bibfnamefont{L.}~\bibnamefont{Krusin-{E}lbaum}},
  \bibinfo{author}{\bibfnamefont{D.~K.} \bibnamefont{Christen}},
  \bibinfo{author}{\bibfnamefont{K.~J.} \bibnamefont{Song}},
  \bibinfo{author}{\bibfnamefont{M.}~\bibnamefont{Paranthaman}},
  \bibinfo{author}{\bibfnamefont{J.~L.} \bibnamefont{Ullmann}},
  \bibinfo{author}{\bibfnamefont{J.~Z.} \bibnamefont{Wu}},
  \bibinfo{author}{\bibfnamefont{Z.~F.} \bibnamefont{Ren}},
  \bibinfo{author}{\bibfnamefont{J.~H.} \bibnamefont{Wang}},
  \bibinfo{author}{\bibfnamefont{J.~E.} \bibnamefont{Tkaczyk}},
  \bibnamefont{et~al.}, \bibinfo{journal}{Appl. Phys. Lett.}
  \textbf{\bibinfo{volume}{71}}, \bibinfo{pages}{536}
  (\bibinfo{year}{1997}{\natexlab{a}}).

\bibitem[{\citenamefont{Schultz et~al.}(1998)\citenamefont{Schultz, Klein,
  Weber, Moss, Zeng, Dou, Sawh, Ren, and Weinstein}}]{Schultz98}
\bibinfo{author}{\bibfnamefont{G.~W.} \bibnamefont{Schultz}},
  \bibinfo{author}{\bibfnamefont{C.}~\bibnamefont{Klein}},
  \bibinfo{author}{\bibfnamefont{H.~W.} \bibnamefont{Weber}},
  \bibinfo{author}{\bibfnamefont{B.}~\bibnamefont{Moss}},
  \bibinfo{author}{\bibfnamefont{R.}~\bibnamefont{Zeng}},
  \bibinfo{author}{\bibfnamefont{S.~X.} \bibnamefont{Dou}},
  \bibinfo{author}{\bibfnamefont{R.}~\bibnamefont{Sawh}},
  \bibinfo{author}{\bibfnamefont{Y.}~\bibnamefont{Ren}}, \bibnamefont{and}
  \bibinfo{author}{\bibfnamefont{R.}~\bibnamefont{Weinstein}},
  \bibinfo{journal}{Appl. Phys. Lett.} \textbf{\bibinfo{volume}{73}},
  \bibinfo{pages}{3935} (\bibinfo{year}{1998}), \bibinfo{note}{and references
  therein.}

\bibitem[{\citenamefont{Thompson et~al.}(1999)\citenamefont{Thompson, Ossandon,
  Krusin-{E}lbaum, Song, Christen, and Ullmann}}]{Thompson99}
\bibinfo{author}{\bibfnamefont{J.~R.} \bibnamefont{Thompson}},
  \bibinfo{author}{\bibfnamefont{J.~G.} \bibnamefont{Ossandon}},
  \bibinfo{author}{\bibfnamefont{L.}~\bibnamefont{Krusin-{E}lbaum}},
  \bibinfo{author}{\bibfnamefont{K.~J.} \bibnamefont{Song}},
  \bibinfo{author}{\bibfnamefont{D.~K.} \bibnamefont{Christen}},
  \bibnamefont{and} \bibinfo{author}{\bibfnamefont{J.~L.}
  \bibnamefont{Ullmann}}, \bibinfo{journal}{Appl. Phys. Lett.}
  \textbf{\bibinfo{volume}{74}}, \bibinfo{pages}{3699} (\bibinfo{year}{1999}).

\bibitem[{\citenamefont{Krusin-{E}lbaum
  et~al.}(1998)\citenamefont{Krusin-{E}lbaum, Blatter, Thompson, Petrov,
  Wheeler, Ullmann, and Chu}}]{LKE98}
\bibinfo{author}{\bibfnamefont{L.}~\bibnamefont{Krusin-{E}lbaum}},
  \bibinfo{author}{\bibfnamefont{G.}~\bibnamefont{Blatter}},
  \bibinfo{author}{\bibfnamefont{J.~R.} \bibnamefont{Thompson}},
  \bibinfo{author}{\bibfnamefont{D.~K.} \bibnamefont{Petrov}},
  \bibinfo{author}{\bibfnamefont{R.}~\bibnamefont{Wheeler}},
  \bibinfo{author}{\bibfnamefont{J.}~\bibnamefont{Ullmann}}, \bibnamefont{and}
  \bibinfo{author}{\bibfnamefont{C.~W.} \bibnamefont{Chu}},
  \bibinfo{journal}{Phys. Rev. Lett.} \textbf{\bibinfo{volume}{81}},
  \bibinfo{pages}{3948} (\bibinfo{year}{1998}).

\bibitem[{\citenamefont{Landau and Ott}(2002)}]{LandauOtt02}
\bibinfo{author}{\bibfnamefont{I.~L.} \bibnamefont{Landau}} \bibnamefont{and}
  \bibinfo{author}{\bibfnamefont{H.~R.} \bibnamefont{Ott}},
  \bibinfo{journal}{Phys. Rev. B} \textbf{\bibinfo{volume}{66}},
  \bibinfo{pages}{144506} (\bibinfo{year}{2002}).

\bibitem[{\citenamefont{Landau and Ott}(2003)}]{LandauOtt03}
\bibinfo{author}{\bibfnamefont{I.~L.} \bibnamefont{Landau}} \bibnamefont{and}
  \bibinfo{author}{\bibfnamefont{H.~R.} \bibnamefont{Ott}},
  \bibinfo{journal}{Physica C} \textbf{\bibinfo{volume}{385}},
  \bibinfo{pages}{544}  (\bibinfo{year}{2003}).

\bibitem[{\citenamefont{Wahl et~al.}(1995)\citenamefont{Wahl, Hardy, Provost,
  Simon, and Buzdin}}]{Wahl95}
\bibinfo{author}{\bibfnamefont{A.}~\bibnamefont{Wahl}},
  \bibinfo{author}{\bibfnamefont{V.}~\bibnamefont{Hardy}},
  \bibinfo{author}{\bibfnamefont{J.}~\bibnamefont{Provost}},
  \bibinfo{author}{\bibfnamefont{C.}~\bibnamefont{Simon}}, \bibnamefont{and}
  \bibinfo{author}{\bibfnamefont{A.}~\bibnamefont{Buzdin}},
  \bibinfo{journal}{Physica C} \textbf{\bibinfo{volume}{250}},
  \bibinfo{pages}{163} (\bibinfo{year}{1995}).

\bibitem[{\citenamefont{Krusin-{E}lbaum
  et~al.}(1997)\citenamefont{Krusin-{E}lbaum, Lopez, Thompson, Wheeler,
  Ullmann, Chu, and Lin}}]{LKE97}
\bibinfo{author}{\bibfnamefont{L.}~\bibnamefont{Krusin-{E}lbaum}},
  \bibinfo{author}{\bibfnamefont{D.}~\bibnamefont{Lopez}},
  \bibinfo{author}{\bibfnamefont{J.~R.} \bibnamefont{Thompson}},
  \bibinfo{author}{\bibfnamefont{R.}~\bibnamefont{Wheeler}},
  \bibinfo{author}{\bibfnamefont{J.}~\bibnamefont{Ullmann}},
  \bibinfo{author}{\bibfnamefont{C.~W.} \bibnamefont{Chu}}, \bibnamefont{and}
  \bibinfo{author}{\bibfnamefont{Q.~M.} \bibnamefont{Lin}},
  \bibinfo{journal}{Nature} \textbf{\bibinfo{volume}{250}},
  \bibinfo{pages}{243} (\bibinfo{year}{1997}).

\bibitem[{\citenamefont{Hao et~al.}(1991)\citenamefont{Hao, Clem, Mc{E}lfresh,
  Civale, Malozemoff, and Holtzberg}}]{HaoClem}
\bibinfo{author}{\bibfnamefont{Z.}~\bibnamefont{Hao}},
  \bibinfo{author}{\bibfnamefont{J.~R.} \bibnamefont{Clem}},
  \bibinfo{author}{\bibfnamefont{M.~W.} \bibnamefont{Mc{E}lfresh}},
  \bibinfo{author}{\bibfnamefont{L.}~\bibnamefont{Civale}},
  \bibinfo{author}{\bibfnamefont{A.~P.} \bibnamefont{Malozemoff}},
  \bibnamefont{and}
  \bibinfo{author}{\bibfnamefont{F.}~\bibnamefont{Holtzberg}},
  \bibinfo{journal}{Phys. Rev. B} \textbf{\bibinfo{volume}{43}},
  \bibinfo{pages}{2844} (\bibinfo{year}{1991}).

\bibitem[{\citenamefont{Kogan et~al.}(1988)\citenamefont{Kogan, Fang, and
  Mitra}}]{Kogan88}
\bibinfo{author}{\bibfnamefont{V.~G.} \bibnamefont{Kogan}},
  \bibinfo{author}{\bibfnamefont{M.~M.} \bibnamefont{Fang}}, \bibnamefont{and}
  \bibinfo{author}{\bibfnamefont{Sreeparna}~\bibnamefont{Mitra}},
  \bibinfo{journal}{Phys. Rev. B} \textbf{\bibinfo{volume}{38}},
  \bibinfo{pages}{11958} (\bibinfo{year}{1988}).

\bibitem[{\citenamefont{Blatter et~al.}(1992)\citenamefont{Blatter,
  Geshkenbein, and Larkin}}]{Blatter92}
\bibinfo{author}{\bibfnamefont{G.}~\bibnamefont{Blatter}},
  \bibinfo{author}{\bibfnamefont{V.~B.} \bibnamefont{Geshkenbein}},
  \bibnamefont{and} \bibinfo{author}{\bibfnamefont{A.~I.}
  \bibnamefont{Larkin}}, \bibinfo{journal}{Phys. Rev. Lett.}
  \textbf{\bibinfo{volume}{68}}, \bibinfo{pages}{875} (\bibinfo{year}{1992}).

\bibitem[{\citenamefont{Li et~al.}(1996)\citenamefont{Li, Fukumoto, Zhu,
  Suenaga, Kaneko, Sata, and Simon}}]{Li96}
\bibinfo{author}{\bibfnamefont{Q.}~\bibnamefont{Li}},
  \bibinfo{author}{\bibfnamefont{Y.}~\bibnamefont{Fukumoto}},
  \bibinfo{author}{\bibfnamefont{Y.}~\bibnamefont{Zhu}},
  \bibinfo{author}{\bibfnamefont{M.}~\bibnamefont{Suenaga}},
  \bibinfo{author}{\bibfnamefont{T.}~\bibnamefont{Kaneko}},
  \bibinfo{author}{\bibfnamefont{K.}~\bibnamefont{Sato}}, \bibnamefont{and}
  \bibinfo{author}{\bibfnamefont{C.}~\bibnamefont{Simon}},
  \bibinfo{journal}{Phys. Rev. B} \textbf{\bibinfo{volume}{54}},
  \bibinfo{pages}{R788} (\bibinfo{year}{1996}).

\bibitem[{\citenamefont{{van der Beek} et~al.}(1996)\citenamefont{{van der
  Beek}, Konczykowski, Li, Kes, and Benoit}}]{vanderBeek96}
\bibinfo{author}{\bibfnamefont{C.~J.} \bibnamefont{{van der Beek}}},
  \bibinfo{author}{\bibfnamefont{M.}~\bibnamefont{Konczykowski}},
  \bibinfo{author}{\bibfnamefont{T.~W.} \bibnamefont{Li}},
  \bibinfo{author}{\bibfnamefont{P.~H.} \bibnamefont{Kes}}, \bibnamefont{and}
  \bibinfo{author}{\bibfnamefont{W.}~\bibnamefont{Benoit}},
  \bibinfo{journal}{Phys. Rev. B} \textbf{\bibinfo{volume}{54}},
  \bibinfo{pages}{R792} (\bibinfo{year}{1996}).

\bibitem[{\citenamefont{{van der Beek} et~al.}(2002)\citenamefont{{van der
  Beek}, Konczykowski, Drost, Kes, Chikumoto, and Bouffard}}]{vanderBeek00}
\bibinfo{author}{\bibfnamefont{C.~J.} \bibnamefont{{van der Beek}}},
  \bibinfo{author}{\bibfnamefont{M.}~\bibnamefont{Konczykowski}},
  \bibinfo{author}{\bibfnamefont{R.~J.} \bibnamefont{Drost}},
  \bibinfo{author}{\bibfnamefont{P.~H.} \bibnamefont{Kes}},
  \bibinfo{author}{\bibfnamefont{N.}~\bibnamefont{Chikumoto}},
  \bibnamefont{and} \bibinfo{author}{\bibfnamefont{S.}~\bibnamefont{Bouffard}},
  \bibinfo{journal}{Phys. Rev. B} \textbf{\bibinfo{volume}{61}},
  \bibinfo{pages}{4259} (\bibinfo{year}{2002}), \bibinfo{note}{and references
  therein}.

\bibitem[{\citenamefont{Thompson
  et~al.}(1997{\natexlab{b}})\citenamefont{Thompson, Krusin-{E}lbaum, Civale,
  Blatter, and Feild}}]{ThompsonPRL97}
\bibinfo{author}{\bibfnamefont{J.~R.} \bibnamefont{Thompson}},
  \bibinfo{author}{\bibfnamefont{L.}~\bibnamefont{Krusin-{E}lbaum}},
  \bibinfo{author}{\bibfnamefont{L.}~\bibnamefont{Civale}},
  \bibinfo{author}{\bibfnamefont{G.}~\bibnamefont{Blatter}}, \bibnamefont{and}
  \bibinfo{author}{\bibfnamefont{C.}~\bibnamefont{Feild}},
  \bibinfo{journal}{Phys. Rev. Lett.} \textbf{\bibinfo{volume}{78}},
  \bibinfo{pages}{3181} (\bibinfo{year}{1997}{\natexlab{b}}).

\bibitem[{\citenamefont{Ossandon et~al.}(2001)\citenamefont{Ossandon, Thompson,
  Krusin-{E}lbaum, Kim, Christen, Song, and Ullmann}}]{Ossandon01}
\bibinfo{author}{\bibfnamefont{J.~G.} \bibnamefont{Ossandon}},
  \bibinfo{author}{\bibfnamefont{J.~R.} \bibnamefont{Thompson}},
  \bibinfo{author}{\bibfnamefont{L.}~\bibnamefont{Krusin-{E}lbaum}},
  \bibinfo{author}{\bibfnamefont{H.~J.} \bibnamefont{Kim}},
  \bibinfo{author}{\bibfnamefont{D.~K.} \bibnamefont{Christen}},
  \bibinfo{author}{\bibfnamefont{K.~J.} \bibnamefont{Song}}, \bibnamefont{and}
  \bibinfo{author}{\bibfnamefont{J.~L.} \bibnamefont{Ullmann}},
  \bibinfo{journal}{Supercond. Sci. Technol.} \textbf{\bibinfo{volume}{14}},
  \bibinfo{pages}{666} (\bibinfo{year}{2001}).

\bibitem[{\citenamefont{Thompson}(1998)}]{ThompsonReview98}
\bibinfo{author}{\bibfnamefont{J.~R.} \bibnamefont{Thompson}}, in
  \emph{\bibinfo{booktitle}{Studies of High Temperature Superconductors}},
  edited by \bibinfo{editor}{\bibfnamefont{A.~V.} \bibnamefont{Narlikar}}
  (\bibinfo{publisher}{Nova Science Publishers}, \bibinfo{address}{Commack,
  NY}, \bibinfo{year}{1998}), vol.~\bibinfo{volume}{26}, pp.
  \bibinfo{pages}{113--131}.

\bibitem[{\citenamefont{Orlando et~al.}(1979)\citenamefont{Orlando, Mc{N}iff,
  Foner, and Beasley}}]{Orlando79}
\bibinfo{author}{\bibfnamefont{T.~P.} \bibnamefont{Orlando}},
  \bibinfo{author}{\bibfnamefont{E.~J.} \bibnamefont{Mc{N}iff},
  \bibfnamefont{Jr.}}, \bibinfo{author}{\bibfnamefont{S.}~\bibnamefont{Foner}},
  \bibnamefont{and} \bibinfo{author}{\bibfnamefont{M.~R.}
  \bibnamefont{Beasley}}, \bibinfo{journal}{Phys. Rev. B}
  \textbf{\bibinfo{volume}{19}}, \bibinfo{pages}{4545} (\bibinfo{year}{1979}).

\bibitem[{\citenamefont{Zhu et~al.}(1993)\citenamefont{Zhu, Cai, Budhani,
  Suenaga, and Welch}}]{Zhu93}
\bibinfo{author}{\bibfnamefont{Y.}~\bibnamefont{Zhu}},
  \bibinfo{author}{\bibfnamefont{Z.~X.} \bibnamefont{Cai}},
  \bibinfo{author}{\bibfnamefont{R.~C.} \bibnamefont{Budhani}},
  \bibinfo{author}{\bibfnamefont{M.}~\bibnamefont{Suenaga}}, \bibnamefont{and}
  \bibinfo{author}{\bibfnamefont{D.~O.} \bibnamefont{Welch}},
  \bibinfo{journal}{Phys. Rev. B} \textbf{\bibinfo{volume}{48}},
  \bibinfo{pages}{6436} (\bibinfo{year}{1993}).

\bibitem[{\citenamefont{Sun et~al.}(1994)\citenamefont{Sun, Thompson, Kerchner,
  Christen, Paranthaman, and Brynestad}}]{Sun94}
\bibinfo{author}{\bibfnamefont{Y.~R.} \bibnamefont{Sun}},
  \bibinfo{author}{\bibfnamefont{J.~R.} \bibnamefont{Thompson}},
  \bibinfo{author}{\bibfnamefont{H.~R.} \bibnamefont{Kerchner}},
  \bibinfo{author}{\bibfnamefont{D.~K.} \bibnamefont{Christen}},
  \bibinfo{author}{\bibfnamefont{M.}~\bibnamefont{Paranthaman}},
  \bibnamefont{and}
  \bibinfo{author}{\bibfnamefont{J.}~\bibnamefont{Brynestad}},
  \bibinfo{journal}{Phys. Rev. B} \textbf{\bibinfo{volume}{50}},
  \bibinfo{pages}{3330} (\bibinfo{year}{1994}).

\bibitem[{\citenamefont{Lewis et~al.}(1995)\citenamefont{Lewis, Vinokur,
  Wagner, and Hinks}}]{Lewis95}
\bibinfo{author}{\bibfnamefont{J.~A.} \bibnamefont{Lewis}},
  \bibinfo{author}{\bibfnamefont{V.~M.} \bibnamefont{Vinokur}},
  \bibinfo{author}{\bibfnamefont{J.}~\bibnamefont{Wagner}}, \bibnamefont{and}
  \bibinfo{author}{\bibfnamefont{D.}~\bibnamefont{Hinks}},
  \bibinfo{journal}{Phys. Rev. B} \textbf{\bibinfo{volume}{52}},
  \bibinfo{pages}{R3852} (\bibinfo{year}{1995}).

\bibitem[{\citenamefont{Kim et~al.}(1995)\citenamefont{Kim, Thompson, Christen,
  Sun, Paranthaman, and Specht}}]{Kim95}
\bibinfo{author}{\bibfnamefont{Y.~C.} \bibnamefont{Kim}},
  \bibinfo{author}{\bibfnamefont{J.~R.} \bibnamefont{Thompson}},
  \bibinfo{author}{\bibfnamefont{D.~K.} \bibnamefont{Christen}},
  \bibinfo{author}{\bibfnamefont{Y.~R.} \bibnamefont{Sun}},
  \bibinfo{author}{\bibfnamefont{M.}~\bibnamefont{Paranthaman}},
  \bibnamefont{and} \bibinfo{author}{\bibfnamefont{E.~D.}
  \bibnamefont{Specht}}, \bibinfo{journal}{Phys. Rev. B}
  \textbf{\bibinfo{volume}{52}}, \bibinfo{pages}{4438} (\bibinfo{year}{1995}).

\bibitem[{\citenamefont{Koshelev and Vinokur}(2001)}]{Koshelev01}
\bibinfo{author}{\bibfnamefont{A.~E.} \bibnamefont{Koshelev}} \bibnamefont{and}
  \bibinfo{author}{\bibfnamefont{V.~M.} \bibnamefont{Vinokur}},
  \bibinfo{journal}{Phys. Rev. B} \textbf{\bibinfo{volume}{64}},
  \bibinfo{pages}{134518} (\bibinfo{year}{2001}).

\bibitem[{\citenamefont{Ossandon
  et~al.}(1992{\natexlab{a}})\citenamefont{Ossandon, Thompson, Christen, Sales,
  Kerchner, Thomson, Sun, Lay, and Tkaczyk}}]{Ossandon92a}
\bibinfo{author}{\bibfnamefont{J.~G.} \bibnamefont{Ossandon}},
  \bibinfo{author}{\bibfnamefont{J.~R.} \bibnamefont{Thompson}},
  \bibinfo{author}{\bibfnamefont{D.~K.} \bibnamefont{Christen}},
  \bibinfo{author}{\bibfnamefont{B.~C.} \bibnamefont{Sales}},
  \bibinfo{author}{\bibfnamefont{H.~R.} \bibnamefont{Kerchner}},
  \bibinfo{author}{\bibfnamefont{J.~O.} \bibnamefont{Thomson}},
  \bibinfo{author}{\bibfnamefont{Y.~R.} \bibnamefont{Sun}},
  \bibinfo{author}{\bibfnamefont{K.~W.} \bibnamefont{Lay}}, \bibnamefont{and}
  \bibinfo{author}{\bibfnamefont{J.~E.} \bibnamefont{Tkaczyk}},
  \bibinfo{journal}{Phys. Rev. B} \textbf{\bibinfo{volume}{45}},
  \bibinfo{pages}{12534} (\bibinfo{year}{1992}{\natexlab{a}}).

\bibitem[{\citenamefont{Ossandon
  et~al.}(1992{\natexlab{b}})\citenamefont{Ossandon, Thompson, Christen, Sales,
  Sun, and Lay}}]{Ossandon92b}
\bibinfo{author}{\bibfnamefont{J.~G.} \bibnamefont{Ossandon}},
  \bibinfo{author}{\bibfnamefont{J.~R.} \bibnamefont{Thompson}},
  \bibinfo{author}{\bibfnamefont{D.~K.} \bibnamefont{Christen}},
  \bibinfo{author}{\bibfnamefont{B.~C.} \bibnamefont{Sales}},
  \bibinfo{author}{\bibfnamefont{Y.~R.} \bibnamefont{Sun}}, \bibnamefont{and}
  \bibinfo{author}{\bibfnamefont{K.~W.} \bibnamefont{Lay}},
  \bibinfo{journal}{Phys. Rev. B} \textbf{\bibinfo{volume}{46}},
  \bibinfo{pages}{3050} (\bibinfo{year}{1992}{\natexlab{b}}).

\end{thebibliography}
\end{document}